\documentclass[12pt]{article}
\newcommand{\bq}{\begin{equation}}
\newcommand{\eq}{\end{equation}}
\newcommand{\ba}{\begin{eqnarray}}
\newcommand{\ea}{\end{eqnarray}}

\newcommand{\nl }{ \nonumber  }

\newcommand{\p}{\partial}
\newcommand{\h}{\hspace{.5cm}}
\newcommand{\s}{\sigma}

\begin{document}
\vspace*{1.cm}
\begin{center}
{\bf NEUMANN AND NEUMANN-ROSOCHATIUS INTEGRABLE SYSTEMS FROM
MEMBRANES ON $AdS_4\times S^7$

\vspace*{1cm} P. Bozhilov}\footnote{e-mail: bozhilov@inrne.bas.bg}

\ \\
\textit{Institute for Nuclear Research and Nuclear Energy,\\
Bulgarian Academy of Sciences, \\ 1784 Sofia, Bulgaria}

\end{center}
\vspace*{.5cm}

It is known that large class of classical string solutions in the
type IIB $AdS_5\times S^5$ background is related to the Neumann
and Neumann-Rosochatius integrable systems, including spiky
strings and giant magnons. It is also interesting if these
integrable systems can be associated with some membrane
configurations in M-theory. We show here that this is indeed the
case by presenting explicitly several types of membrane embedding
in $AdS_4\times S^7$ with the searched properties.


\vspace*{.5cm} {\bf Keywords:} M-theory, integrable systems,
AdS-CFT duality.

\section{Introduction}
The AdS/CFT correspondence predicts that the string theory on
$AdS_5\times S^5$ should be dual to $\mathcal{N}=4$ SYM theory in
four dimensions \cite{M97}, \cite{GKP98}, \cite{W98}. The spectrum
of the string states and of the operators in SYM should be the
same. The recent checks of this conjecture {\it beyond} the
supergravity approximation are connected to the idea to search for
string solutions, which in the semiclassical limit are related to
the anomalous dimensions of certain gauge invariant operators in
the planar limit of SYM \cite{BMN02}, \cite{GKP02}. On the field
theory side, it was found that the corresponding dilatation
operator is connected to the Hamiltonian of integrable Heisenberg
spin chain \cite{MZ}. On the string side, it was established that
large set of classical string solutions follow from embeddings ,
which reduce the solution of the string equations of motion and
constraints to the study of the Neumann and Neumann-Rosochatius
integrable systems in the presence of conformal gauge constraints
\cite{N}, \cite{NR1}, \cite{NR2}.

In \cite{N} it was shown that solitonic solutions of the classical
string action on the type IIB $AdS_5\times S^5$ background that
carry charges of the Cartan subalgebra of the global symmetry
group can be classified in terms of periodic solutions of the
Neumann dynamical system \cite{Neumann}, which is Liouville
integrable \cite{Per}. A particular string solution was also
identified, whose classical energy reproduces {\it exactly} the
one-loop anomalous dimension of a certain set of SYM operators
with two independent R-charges.

A general class of rotating closed string solutions in
$AdS_5\times S^5$ was shown to be connected to the
Neumann-Rosochatius integrable system \cite{NR} in \cite{NR1}.

Let us note that the first multi-spin string solutions were found
in \cite{hep-th/0304255} and \cite{hep-th/0306143}, where it was
pointed out that the classical energy of the strings admitted a
perturbative expansion in powers of the 't Hooft coupling
$\lambda$, and therefore could be compared with perturbative field
theory computations. Actually, the very first two-spin string was
found in \cite{hep-th/0204226} but its importance was not
understood at that time. Papers \cite{N} and \cite{NR1}
generalized these multi-spin solutions.

The first perturbative field theory computation of anomalous
dimensions of long SYM operators was done in
\cite{hep-th/0306139}, where the operators dual to strings from
\cite{hep-th/0304255} and \cite{hep-th/0306143} were identified
and exact one-loop matching was shown.

It was found in \cite{NR2} that, working in conformal gauge, the
spiky strings \cite{MK,SR} and giant magnons \cite{HM}-
\cite{PS07} can be also accommodated by a version of the
Neumann-Rosochatius system. The authors of \cite{NR2} was able to
describe in detail a giant magnon solution with two additional
angular momenta and to show that it can be interpreted as a
superposition of two magnons moving with the same speed. In
addition, they considered the spin chain side and described the
corresponding state as that of two bound states in the infinite
$SU(3)$ spin chain. The Bethe ansatz wave function for such bound
state was also constructed.

It was also shown recently that magnon-like dispersion relations
can arise from M-theory \cite{BR06}, \cite{B06}. That is why, it
is interesting if the Neumann and Neumann-Rosochatius integrable
systems can be associated with some M2-brane configurations. In
this paper, we prove that this is indeed the case by presenting
explicitly several types of membrane embedding in $AdS_4\times
S^7$ with the desired properties.

\setcounter{equation}{0}
\section{Short review of the string case}
Our aim here is to briefly describe part of the results obtained
in \cite{N}, \cite{NR1} and \cite{NR2}, concerning the
correspondence between different type of string solutions on
$AdS_5\times S^5$ in conformal gauge with the Neumann and
Neumann-Rosochatius like integrable systems. Then we show how to
generalize these results to the case of diagonal worldsheet gauge.

The action for the bosonic part of the classical closed string
moving in the $AdS_5\times S^5$ background, in conformal gauge,
can be written as\footnote{We follow the notation of \cite{N}.}
\ba\label{sa1} I=-\frac{\sqrt{\lambda}}{4\pi}\int d\tau d\sigma
\left[G_{mn}^{\left(AdS_5\right)}(x)\p_a x^m\p^a x^n +
G_{pq}^{\left(S^5\right)}(y)\p_a y^p\p^a y^q\right],\h
\sqrt{\lambda}=2\pi R^2T,\ea where the two metrics are given by
\ba\label{ma} &&\left(ds^2\right)_{AdS_5}=-\cosh^2\rho
dt^2+d\rho^2+\sinh^2\rho\left(d\theta^2+\sin^2\theta
d\phi^2+\cos^2\theta d\varphi^2\right), \\ \label{ms}
&&\left(ds^2\right)_{S_5}=d\gamma^2+\cos^2\gamma
d\varphi_3^2+\sin^2\gamma\left(d\psi^2+\cos^2\psi d\varphi_1^2
+\sin^2\psi d\varphi_2^2\right).\ea

The action (\ref{sa1}) can be represented as action for the
$O(6)\times SO(4,2)$ sigma-model \ba\label{sma1}
I=\frac{\sqrt{\lambda}}{4\pi}\int d\tau d\sigma
\left(L_S+L_{AdS}\right),\ea where \ba\label{ls}
&&L_S=-\frac{1}{2}\p_a
X_M\p^aX_M+\frac{1}{2}\Lambda\left(X_MX_M-1\right),\h M=1,...,6,
\\ \label{la}
&&L_{AdS}=-\frac{1}{2}\eta_{MN}\p_aY_M\p^aY_N +
\frac{1}{2}\tilde{\Lambda}\left(\eta_{MN}Y_MY_N+1\right),\\ \nl
&&M=0,...,5,\h \eta_{MN}=(-1,1,1,1,1,-1).\ea The embedding
coordinates $X_M$, $Y_M$ are related to the ones in (\ref{ma}),
(\ref{ms}) as follows \ba\label{ex} X_1+iX_2=\sin\gamma\cos\psi
e^{i\varphi_1},\h X_3+iX_4=\sin\gamma\sin\psi e^{i\varphi_2},\h
X_5+iX_6=\cos\gamma e^{i\varphi_3},\\ \label{ey}
Y_1+iY_2=\sinh\rho\sin\theta e^{i\phi},\h
Y_3+iY_4=\sinh\rho\cos\theta e^{i\varphi},\h Y_5+iY_0=\cosh\rho
e^{it}.\ea The action (\ref{sma1}) must be supplemented with the
two conformal gauge constraints.

Further on, the following ansatz for the string embedding has been
proposed in \cite{N} \ba\label{se1} &&Y_1,...,Y_4=0,\h
Y_5+iY_0=e^{i\kappa\tau}, \\ \nl &&X_1+iX_2=x_1(\s)
e^{i\omega_1\tau},\h X_3+iX_4=x_2(\s) e^{i\omega_2\tau},\h
X_5+iX_6=x_3(\s) e^{i\omega_3\tau}.\ea It corresponds to string
located at the center of $AdS_5$ and rotating in $S^5$. Replacing
(\ref{se1}) into (\ref{ls}), (\ref{la}), one obtains the string
Lagrangian (prime is used for $d/d\s$) \ba\nl
L_S+L_{AdS}=-\frac{1}{2}\left[\sum_{i=1}^{3}\left(x_i'^2-\omega_i^2x_i^2\right)
+\kappa^2\right]+
\frac{1}{2}\Lambda\left(\sum_{i=1}^{3}x_i^2-1\right).\ea After
changing the overall sign and neglecting the constant term as in
\cite{N}, one arrives at \ba\label{N} L
=\frac{1}{2}\sum_{i=1}^{3}\left(x_i'^2-\omega_i^2x_i^2\right) +
\frac{1}{2}\Lambda\left(\sum_{i=1}^{3}x_i^2-1\right).\ea $L$
describes three dimensional harmonic oscillator constrained to
remain on a unit two-sphere. This is particular case of the
$n$-dimensional Neumann dynamical system \cite{Neumann}, which is
Liouville {\it integrable} \cite{Per}. In the case under
consideration, the only nontrivial Virasoro constraint implies
that the energy $H$ of the Neumann system is given by
\ba\label{NE} H
=\frac{1}{2}\sum_{i=1}^{3}\left(x_i'^2+\omega_i^2x_i^2\right)=
\frac{1}{2}\kappa^2.\ea

In order to obtain the relevant closed string solutions, we should
impose periodicity conditions on $x_i$: \ba\nl
x_i(\s)=x_i(\s+2\pi).\ea

Another string embedding is possible, related to Neumann like
integrable system \cite{N} \ba\label{se2} Y_1+iY_2=y_1(\s)
e^{i\omega_1\tau},\h Y_3+iY_4=y_2(\s) e^{i\omega_2\tau},\h
Y_5+iY_0=y_3(\s) e^{i\omega_3\tau}.\ea It corresponds to
multi-spin strings rotating not in $S^5$ but in $AdS_5$ instead.
Now $t=\omega_3\tau$, so the equality $\omega_3=\kappa$ holds. The
relevant effective mechanical system describing this class of
rotating solutions has the following Lagrangian
\ba\label{Nl}\tilde{L}=\frac{1}{2}\eta_{ij}\left(y_i'y_j'-\omega_i^2y_iy_j\right)
+ \frac{1}{2}\tilde{\Lambda}\left(\eta_{ij}y_iy_j-1\right), \h
\eta_{ij}=diag(-1,-1,1).\ea Comparing this with the Neumann
Lagrangian (\ref{N}), one concludes that (\ref{Nl}) corresponds to
a system, which is similar to the Neumann integrable system, but
with indefinite signature - $\delta_{ij}$ replaced by $\eta_{ij}$.
The relation to the $S^5$ case is through the analytic
continuation \ba\nl x_1\to iy_1,\h x_2\to iy_2.\ea

The results presented above have been generalized in \cite{NR1} to
correspondence between closed strings in $AdS_5\times S^5$ and the
Neumann-Rosochatius integrable system \cite{NR}. This has been
achieved by using more general ansatz for the string embedding.
Two such types of embedding have been given in \cite{NR1}. The
first one is\footnote{We follow the notation of \cite{NR1}.}
\ba\label{se3} &&Y_1,...,Y_4=0,\h Y_5+iY_0=e^{i\kappa\tau}, \\ \nl
&&X_1+iX_2=r_1(\s) e^{i\left[\omega_1\tau+\alpha_1(\s)\right]},\\
\nl
&&X_3+iX_4=r_2(\s)e^{i\left[(\omega_2\tau+\alpha_2(\s)\right]},\\
\nl &&X_5+iX_6=r_3(\s)
e^{i\left[\omega_3\tau+\alpha_3(\s)\right]}.\ea To find the
corresponding closed string solutions, one imposes the periodicity
conditions \ba\nl r_i(\s+2\pi)=r_i(\s),\h
\alpha_i(\s+2\pi)=\alpha_i+2\pi m_i,\h m_i=0,\pm 1,\pm 2,... .\ea

The ansatz (\ref{se3}) leads to the following Lagrangian
\ba\label{NRa} L =\frac{1}{2}\sum_{i=1}^{3}\left(r_i'^2 +
r_i^2\alpha_i'^2-\omega_i^2r_i^2\right) -
\frac{1}{2}\Lambda\left(\sum_{i=1}^{3}r_i^2-1\right).\ea The
equations of motion for the variables $\alpha_i(\s)$ can be easily
integrated once \ba\label{a'} \alpha_i'=\frac{v_i}{r_i^2},\h
v_i=constants.\ea Substituting (\ref{a'}) back into (\ref{NRa}),
one receives an effective Lagrangian for the three real
coordinates $r_i(\s)$\footnote{Following \cite{NR1}, we change the
signs of the terms $\sim \alpha_i'^2$.} \ba\label{NR} L
=\frac{1}{2}\sum_{i=1}^{3}\left(r_i'^2 -\omega_i^2r_i^2
-\frac{v_i^2}{r_i^2}\right) -
\frac{1}{2}\Lambda\left(\sum_{i=1}^{3}r_i^2-1\right).\ea When
$\alpha_i$ are constants, i.e. $v_i=0$, (\ref{NR}) reduces to the
Neumann Lagrangian (\ref{N}). For non-zero $v_i$, the Lagrangian
(\ref{NR}) describes the Neumann-Rosochatius integrable system.
The Virasoro constraints take the form \ba\nl
&&\sum_{i=1}^{3}\left(r_i'^2 +\omega_i^2r_i^2
+\frac{v_i^2}{r_i^2}\right)=\kappa^2,\\ \nl
&&\sum_{i=1}^{3}\omega_iv_i=0.\ea As a consequence of the second
equality, only two of the three integrals of motion $v_i$ are
independent of $\omega_i$.

The second type of embedding proposed in \cite{NR1} is for the
case when the string rotates in both $AdS_5$ and $S^5$. It is
given by (\ref{se3}) for $X_1,...,X_6$ and \ba\label{se4}
&&Y_5+iY_0=\mbox{r}_0(\s)e^{i\left[w_0\tau+\beta_0(\s)\right]},\\
\nl
&&Y_1+iY_2=\mbox{r}_1(\s)e^{i\left[w_1\tau+\beta_1(\s)\right]},
\\ \nl
&&Y_3+iY_4=\mbox{r}_2(\s)e^{i\left[w_2\tau+\beta_2(\s)\right]}.\ea
To satisfy the closed string periodicity conditions, one needs the
following equalities to hold ($k_r$ are integers) \ba\nl
\mbox{r}_r(\s+2\pi)=\mbox{r}_r(\s),\h
\beta_r(\s+2\pi)=\beta_r(\s)+2\pi k_r,\h r=0,1,2.\ea Requiring the
time coordinate to be single-valued (considering a universal cover
of $AdS_5$), i.e. ignoring windings in time direction, and also
renaming $w_0$ to $\kappa$, one obtains \ba\nl k_0=0,\h w_0\equiv
\kappa.\ea

The mechanical system corresponding to the above embedding is
described by the sum of the Lagrangian (\ref{NR}) and the
following one \ba\label{NRL}
\tilde{L}=\frac{1}{2}\eta^{rs}\left(\mbox{r}'_r\mbox{r}'_s
-w_r^2\mbox{r}_s\mbox{r}_s -
\frac{u_ru_s}{\mbox{r}_r\mbox{r}_s}\right)
-\frac{1}{2}\tilde{\Lambda}\left(\eta^{rs}
\mbox{r}_r\mbox{r}_s+1\right),\h \eta^{rs}=(-1,1,1),\ea which
represents an integrable system too.

For the present case, the equations of motion for $r_i$ and
$\mbox{r}_s$, following from (\ref{NR}) and (\ref{NRL})
respectively, decouple. However, in the conformal gauge
constraints, the variables of the two Neumann-Rosochatius systems
are mixed. More precisely, the Virasoro constraints now read
\ba\nl
&&\mbox{r}'^2_0+\kappa^2\mbox{r}_0^2+\frac{u_0^2}{\mbox{r}_0^2}=
\sum_{a=1}^{2}\left(\mbox{r}_a'^2 +w_a^2\mbox{r}_a^2
+\frac{u_a^2}{\mbox{r}_a^2}\right)+\sum_{i=1}^{3}\left(r_i'^2
+\omega_i^2r_i^2 +\frac{v_i^2}{r_i^2}\right),
\\ \nl &&\kappa u_0=\sum_{a=1}^{2}w_au_a +
\sum_{i=1}^{3}\omega_iv_i,\ea where \ba\nl
\mbox{r}_0^2-\sum_{a=1}^{2}\mbox{r}_a^2=1,\h
\sum_{i=1}^{3}r_i^2=1.\ea We also require the periodicity
conditions \cite{NR1} \ba\nl
u_s\int_{0}^{2\pi}\frac{d\s}{\mbox{r}_s^2(\s)}=2\pi k_s\ea to be
fulfilled. Then $k_0=0$ implies $u_0=0$ as a consequence of the
single-valuedness of the time coordinate $t$.

The authors of \cite{NR2}, inspired by the recent development in
string/CFT duality, proposed new string embedding, which
incorporates the spiky strings \cite{MK,SR} and giant magnons
\cite{HM}-\cite{WHH06} on $S^5$. They showed that such string
solutions can be also accommodated by a version of the
Neumann-Rosochatius integrable system. The appropriate embedding
is given by \ba\label{se5} &&Y_1,...,Y_4=0,\h
Y_5+iY_0=e^{i\kappa\tau}, \\ \nl
&&X_1+iX_2=r_1(\xi) e^{i\left[\omega_1\tau+\mu_1(\xi)\right]},\\
\nl
&&X_3+iX_4=r_2(\xi)e^{i\left[(\omega_2\tau+\mu_2(\xi)\right]},\\
\nl &&X_5+iX_6=r_3(\xi)
e^{i\left[\omega_3\tau+\mu_3(\xi)\right]},\ea where \ba\nl
\xi=\alpha\sigma+\beta\tau .\ea This ansatz leads to the
Lagrangian \cite{NR2} \ba\label{NRSM}
L=\sum_{i=1}^{3}\left[\left(\alpha^2-\beta^2\right)r_i'^2
-\frac{1}{\alpha^2-\beta^2}\frac{C_i^2}{r_i^2} -
\frac{\alpha^2}{\alpha^2-\beta^2}\omega_i^2 r_i^2\right] +
\Lambda\left(\sum_{i=1}^{3}r_i^2-1\right),\ea which describes the
standard Neumann-Rosochatius integrable system. The corresponding
Hamiltonian is \ba\nl H=
\sum_{i=1}^{3}\left[\left(\alpha^2-\beta^2\right)r_i'^2
+\frac{1}{\alpha^2-\beta^2}\frac{C_i^2}{r_i^2} +
\frac{\alpha^2}{\alpha^2-\beta^2}\omega_i^2 r_i^2\right].\ea The
Virasoro constraints are satisfied if \ba\nl
H=\frac{\alpha^2+\beta^2}{\alpha^2-\beta^2}\kappa^2,\h
\sum_{i=1}^{3}\omega_iC_i + \beta\kappa^2=0.\ea The periodicity
conditions read \ba\nl r_i(\xi+2\pi\alpha)=r_i(\xi),\h
\mu_i(\xi+2\pi\alpha)=\mu_i(\xi)+2\pi n_i,\ea where $n_i$ are
integer winding numbers. The second condition implies \ba\nl
\frac{C_i}{2\pi}\int_{0}^{2\pi\alpha}\frac{d \xi}{r_i^2}=
(\alpha^2-\beta^2)n_i-\alpha\beta\omega_i.\ea Thus the general
solution for the ansatz (\ref{se5}) can be constructed in terms of
the usual solutions of the Neumann-Rosochatius system. There are
five independent integrals of motion, which reduce the equations
of motion to a system of first order differential equations that
can be directly integrated \cite{N}.

All the above results are obtained in conformal gauge. In order to
make connection with the membrane case, we will formulate the
problem in the framework of the more general diagonal worldsheet
gauge. In this gauge, the Polyakov action and constraints are
given by \ba\label{gfsa} &&S_{S}=\int d^{2}\xi \mathcal{L}_{S}=
\int
d^{2}\xi\frac{1}{4\lambda^0}\Bigl[G_{00}-\left(2\lambda^0T\right)^2
G_{11}\Bigr],
\\ \label{gf00} &&G_{00}+\left(2\lambda^0T\right)^2 G_{11}=0,
\\ \label{gf01} &&G_{01}=0,\ea where
\ba\nl &&G_{mn}= g_{MN}\p_m X^M\p_n X^N,\\ \nl
&&\left[\p_m=\p/\p\xi^m, \h m = (0,1),\h
(\xi^0,\xi^1)=(\tau,\sigma),\h M = (0,1,\ldots,9)\right],\ea is
the induced metric and $\lambda^0$ is Lagrange multiplier. The
usually used conformal gauge corresponds to $2\lambda^0T=1$.

The general string embedding in $AdS_5\times S^5$ of the type we
are interested in can be written as \ba\nl
&&Z_{s}=R\mbox{r}_s(\xi^m)e^{i\phi_s(\xi^m)},\h s=(0,1,2),\h
\eta^{rs} \mbox{r}_r\mbox{r}_s+1=0,\h \eta^{rs}=(-1,1,1), \\
\label{gse} &&W_{i}=Rr_i(\xi^m)e^{i\varphi_i(\xi^m)},\h
i=(1,2,3),\h \delta_{ij} r_ir_j-1=0.\ea For this embedding, the
induced metric takes the form \ba\label{fsim}
&&G_{mn}=\eta^{rs}\p_{(m}Z_r\p_{n)}\bar{Z_s} +
\delta_{ij}\p_{(m}W_i\p_{n)}\bar{W_j}= \\ \nl
&&R^2\left[\sum_{r,s=0}^{2}\eta^{rs}\left(\p_m\mbox{r}_r\p_n\mbox{r}_s
+ \mbox{r}_r^2\p_m\phi_r\p_n\phi_s\right) +
\sum_{i=1}^{3}\left(\p_mr_i\p_nr_i +
r_i^2\p_m\varphi_i\p_n\varphi_i\right)\right].\ea The expression
(\ref{fsim}) for $G_{mn}$ must be replaced into (\ref{gfsa}),
(\ref{gf00}) and (\ref{gf01}). Correspondingly, the string
Lagrangian will be \ba\label{fssl} \mathcal{L}= \mathcal{L}_{S}+
\Lambda_A(\eta^{rs} \mbox{r}_r\mbox{r}_s+1) +
\Lambda_S(\delta_{ij} r_ir_j-1),\ea where $\Lambda_A$ and
$\Lambda_S$ are Lagrange multipliers.

As an example, let us choose the following ansatz for the string
embedding of the type (\ref{gse}) \ba\nl Z_0=R e^{i\kappa\tau},\h
Z_1=Z_2=0,\h W_{i}=Rr_i(\sigma)e^{i\omega_i\tau},\ea which implies
\ba\nl \mbox{r}_0=1,\h \mbox{r}_1=\mbox{r}_2=0;\h
\phi_0=\kappa\tau,\h \varphi_i=\omega_i\tau.\ea Then (\ref{fssl})
reduces to (prime is used for $d/d\s$) \ba\nl
\mathcal{L}=-\frac{R^2}{4\lambda^0}\left\{\sum_{i=1}^{3}\left[(2\lambda^0T)^2r_i'^2-\omega_i^2r_i^2\right]
+ \kappa^2\right\} + \Lambda_S\left(\sum_{i=1}^{3}
r_i^2-1\right).\ea After changing the overall sign and neglecting
the constant term as in \cite{N}, one obtains \ba\nl
L=\frac{R^2}{4\lambda^0}\sum_{i=1}^{3}\left[(2\lambda^0T)^2r_i'^2-\omega_i^2r_i^2\right]
 + \Lambda_S\left(\sum_{i=1}^{3} r_i^2-1\right),\ea which in
conformal gauge ($2\lambda^0T=1$) is equivalent to (\ref{N}). The
constraint (\ref{gf00}) gives the corresponding Hamiltonian \ba\nl
H\sim
\sum_{i=1}^{3}\left[(2\lambda^0T)^2r_i'^2+\omega_i^2r_i^2\right]=\kappa^2.\ea
The other constraint (\ref{gf01}) is satisfied identically.

In the same way, one can generalize the other previously obtained
results \cite{N,NR1,NR2} to the case of diagonal worldsheet gauge.

\setcounter{equation}{0}
\section{Membranes on $AdS_4\times S^7$}
Turning to the membrane case, let us first write down the gauge
fixed membrane action and constraints in diagonal worldvolume
gauge, we are going to work with: \ba\label{omagf} &&S_{M}=\int
d^{3}\xi \mathcal{L}_{M}= \int
d^{3}\xi\left\{\frac{1}{4\lambda^0}\Bigl[G_{00}-\left(2\lambda^0T_2\right)^2\det
G_{ij}\Bigr] + T_2 C_{012}\right\},
\\ \label{00gf} &&G_{00}+\left(2\lambda^0T_2\right)^2\det G_{ij}=0,
\\ \label{0igf} &&G_{0i}=0.\ea They {\it coincide} with the
frequently used gauge fixed Polyakov type action and constraints
after the identification $2\lambda^0T_2=L=const$, where
$\lambda^0$ is Lagrange multiplier and $T_2$ is the membrane
tension. In (\ref{omagf})-(\ref{0igf}), the fields induced on the
membrane worldvolume $G_{mn}$ and $C_{012}$ are given by
\ba\label{im} &&G_{mn}= g_{MN}\p_m X^M\p_n X^N,\h C_{012}=
c_{MNP}\p_{0}X^{M}\p_{1}X^{N}\p_{2}X^{P}, \\ \nl
&&\p_m=\p/\p\xi^m,\h m = (0,i) = (0,1,2),\\ \nl
&&(\xi^0,\xi^1,\xi^2)=(\tau,\sigma_1,\sigma_2),\h M =
(0,1,\ldots,10),\ea where $g_{MN}$ and $c_{MNP}$ are the
components of the target space metric and 3-form gauge field
respectively.

Searching for membrane configurations in $AdS_4\times S^7$, which
correspond to the Neumann or Neumann-Rosochatius integrable
systems, we should first eliminate the membrane interaction with
the background 3-form field on $AdS_4$, to ensure more close
analogy with the strings on $AdS_5\times S^5$. To make our choice,
let us write down the background. It can be parameterized as
follows \ba\nl &&ds^2=(2l_p\mathcal{R})^2\left[-\cosh^2\rho
dt^2+d\rho^2+\sinh^2\rho\left(d\alpha^2+\sin^2\alpha
d\beta^2\right) + 4d\Omega_7^2\right], \\ \nl
&&c_{(3)}=(2l_p\mathcal{R})^3\sinh^3\rho\sin\alpha dt\wedge
d\alpha\wedge d\beta.\ea Since we want the membrane to have
nonzero conserved energy and spin on $AdS$, the possible choice,
for which the interaction with the $c_{(3)}$ field disappears, is
to fix the angle $\alpha$\footnote{Of course, we can fix the angle
$\beta$ instead of $\alpha$. Then, in the corresponding subspace
of $AdS_4$, $\alpha$ will be the isometry coordinate associated
with the conserved spin. The difference is that $\beta$ is the
isometry coordinate in the initial $AdS_4$ space.}: \ba\nl
\alpha=\alpha_0=const.\ea The metric of the corresponding subspace
of $AdS_4$ is \ba\label{sub}
&&ds^2_{sub}=(2l_p\mathcal{R})^2\left(-\cosh^2\rho
dt^2+d\rho^2+\sinh^2\rho\sin^2\alpha_0 d\beta^2\right)=
\\ \nl &&(2l_p\mathcal{R})^2\left[-\cosh^2\rho
dt^2+d\rho^2+\sinh^2\rho d(\beta\sin\alpha_0)^2\right].\ea
Therefore, the appropriate membrane embedding into (\ref{sub}) and
$S^7$ is \ba\nl
&&Z_{\mu}=2l_p\mathcal{R}\mbox{r}_\mu(\xi^m)e^{i\phi_\mu(\xi^m)},\h
\mu=(0,1),\h \phi_{\mu}=(\phi_0,\phi_1)=(t,\beta\sin\alpha_0),\\
\label{gme}
&&\h\h\h\h\h\h\h\h\h\h \eta^{\mu\nu} \mbox{r}_\mu\mbox{r}_\nu+1=0,\h \eta^{\mu\nu}=(-1,1), \\
\nl &&W_{a}=4l_p\mathcal{R}r_a(\xi^m)e^{i\varphi_a(\xi^m)},\h
a=(1,2,3,4),\h \delta_{ab} r_ar_b-1=0.\ea For this embedding, the
induced metric is given by \ba\label{mim}
&&G_{mn}=\eta^{\mu\nu}\p_{(m}Z_\mu\p_{n)}\bar{Z_\nu} +
\delta_{ab}\p_{(m}W_a\p_{n)}\bar{W_b}= \\ \nl
&&(2l_p\mathcal{R})^2\left[\sum_{\mu,\nu=0}^{1}\eta^{\mu\nu}\left(\p_m\mbox{r}_\mu\p_n\mbox{r}_\nu
+ \mbox{r}_\mu^2\p_m\phi_\mu\p_n\phi_\nu\right) +
4\sum_{a=1}^{4}\left(\p_mr_a\p_nr_a +
r_a^2\p_m\varphi_a\p_n\varphi_a\right)\right].\ea We will use the
expression (\ref{mim}) for $G_{mn}$ in (\ref{omagf}), (\ref{00gf})
and (\ref{0igf}). Correspondingly, the membrane Lagrangian becomes
\ba\label{geml}
\mathcal{L}=\mathcal{L}_{M}+\Lambda_A(\eta^{\mu\nu}
\mbox{r}_\mu\mbox{r}_\nu+1)+\Lambda_S(\delta_{ab} r_ar_b-1).\ea

\subsection{Membranes and the Neumann system}
Here, we propose two membrane embeddings in $AdS_4\times S^7$
related to the Neumann integrable system.

Let us begin with the following ansatz for the membrane embedding
of the type (\ref{gme})\ba\label{1}
Z_{0}=2l_p\mathcal{R}e^{i\kappa\tau},\h Z_1=0,\h
W_{a}=4l_p\mathcal{R}r_a(\tau)e^{i\omega_{ai}\sigma_i}.\ea This
implies \ba\nl \mbox{r}_0=1,\h \mbox{r}_1=0,\h
\phi_0=\kappa\tau,\h \varphi_a=\omega_{ai}\sigma_i.\ea Then
(\ref{geml}) takes the form (over-dot is used for $d/d\tau$)
\ba\label{NmL1}
\mathcal{L}&=&\frac{(4l_p\mathcal{R})^2}{4\lambda^0}\left[\sum_{a=1}^{4}\dot{r}_a^2
-\left(8\lambda^0T_2l_p\mathcal{R}\right)^2
\sum_{a<b=1}^{4}(\omega_{a1}\omega_{b2}-\omega_{a2}\omega_{b1})^2r_a^2r_b^2
-(\kappa/2)^2\right]\\ \nl
&+&\Lambda_S\left(\sum_{a=1}^{4}r_a^2-1\right).\ea It is clear
that for arbitrary and different values of the winding numbers
$\omega_{ai}$, the potential terms in the above Lagrangian are of
forth order with respect to $r_a$. As far as we are interested in
obtaining membrane configurations with quadratic effective
potential, our proposal is to make the following choice ($a, b,
c\ne 0$ are constants) \ba\label{c1}
&&\omega_{12}=\omega_{22}=\omega_{31}=\omega_{41}=0,\h
\omega_{32}=\pm\omega_{42}=\omega, \\ \nl &&
r_3(\tau)=a\sin(b\tau+c),\h r_4(\tau)=a\cos(b\tau+c),\h a<1.\ea
This reduces the membrane Lagrangian to \ba\nl
\mathcal{L}&=&\frac{(4l_p\mathcal{R})^2}{4\lambda^0}\left[\sum_{a=1}^{2}\dot{r}_a^2
-\left(8\lambda^0T_2l_p\mathcal{R}a\omega\right)^2
\sum_{a=1}^{2}\omega_{a1}^2r_a^2 + (ab)^2 -(\kappa/2)^2\right]\\
\nl &+&\Lambda_S\left(\sum_{a=1}^{2}r_a^2+a^2-1\right).\ea After
neglecting the constat terms here, one arrives at \ba\nl
L=\frac{(4l_p\mathcal{R})^2}{4\lambda^0}\sum_{a=1}^{2}\left[\dot{r}_a^2
-\left(8\lambda^0T_2l_p\mathcal{R}a\omega\right)^2
\omega_{a1}^2r_a^2\right]+\Lambda_S\left[\sum_{a=1}^{2}r_a^2-(1-a^2)\right].\ea
The Lagrangian $L$ describes two-dimensional harmonic oscillator,
constrained to remain on a circle of radius $\sqrt{1-a^2}$.
Obviously, this is particular case of the Neumann integrable
system. The constraint (\ref{00gf}) gives the Hamiltonian
corresponding to $L$ \ba\nl H\sim \sum_{a=1}^{2}\left[\dot{r}_a^2
+\left(8\lambda^0T_2l_p\mathcal{R}a\omega\right)^2
\omega_{a1}^2r_a^2\right]=(\kappa/2)^2-(ab)^2,\ea while the
remaining constraints (\ref{0igf}) are satisfied identically.

The next ansatz for membrane embedding we will consider is
\ba\label{2} Z_{0}=2l_p\mathcal{R}e^{i\kappa\tau},\h Z_1=0,\h
W_{a}=4l_p\mathcal{R}r_a(\sigma_i)e^{i\omega_{a}\tau},\ea for
which (\ref{geml}) reduces to \ba\nl
\mathcal{L}&=&-\frac{(4l_p\mathcal{R})^2}{4\lambda^0}\left[\left(8\lambda^0T_2l_p\mathcal{R}\right)^2
\sum_{a<b=1}^{4}(\p_1r_a\p_2r_b-\p_2r_a\p_1r_b)^2-\sum_{a=1}^{4}\omega_a^2r_a^2
+(\kappa/2)^2\right]\\ \label{NmL2}
&+&\Lambda_S\left(\sum_{a=1}^{4}r_a^2-1\right).\ea Here we have
quadratic potential, but in the general case, the kinetic term is
not of the type we are searching for. To fix the problem, we set
\ba\label{smch} &&r_{1}=r_{1}(\sigma_1),\h
r_{2}=r_{2}(\sigma_1),\h \omega_3=\pm\omega_4=\omega,\\ \nl
&&r_3(\sigma_2)=a\sin(b\sigma_2+c),\h
r_4(\sigma_2)=a\cos(b\sigma_2+c),\h a<1.\ea This leads to the
Lagrangian (prime is used for $d/d\s_1$)\footnote{After changing
the overall sign and neglecting the constant terms.}
\ba\label{nl2}
L=\frac{(4l_p\mathcal{R})^2}{4\lambda^0}\sum_{a=1}^{2}\left[\left(8\lambda^0T_2l_p\mathcal{R}ab\right)^2
r_a'^2-\omega_a^2r_a^2\right] +
\Lambda_S\left[\sum_{a=1}^{2}r_a^2-(1-a^2)\right],\ea which is
already of the Neumann type. The corresponding Hamiltonian is
given by the constraint (\ref{00gf}) \ba\nl H\sim
\sum_{a=1}^{2}\left[\left(8\lambda^0T_2l_p\mathcal{R}ab\right)^2
r_a'^2+\omega_a^2r_a^2\right]=(\kappa/2)^2- (a\omega)^2.\ea The
other two constraints (\ref{0igf}) are satisfied identically.

\subsection{Membranes and the Neumann-Rosochatius system}
In this subsection, we propose three different membrane embeddings
in $AdS_4\times S^7$ of the type (\ref{gme}), which are connected
with particular cases of the Neumann-Rosochatius integrable
system.

The first one is \ba\label{3}
Z_{0}=2l_p\mathcal{R}e^{i\kappa\tau},\h Z_1=0,\h
W_{a}=4l_p\mathcal{R}r_a(\tau)e^{i[\omega_{ai}\sigma_i+\alpha_a(\tau)]}.\ea
It leads to the following membrane Lagrangian \ba\nl
\mathcal{L}&=&\frac{(4l_p\mathcal{R})^2}{4\lambda^0}\left[\sum_{a=1}^{4}
\left(\dot{r}_a^2+ r_a^2\dot{\alpha}_a^2\right)
-\left(8\lambda^0T_2l_p\mathcal{R}\right)^2
\sum_{a<b=1}^{4}(\omega_{a1}\omega_{b2}-\omega_{a2}\omega_{b1})^2r_a^2r_b^2
-(\kappa/2)^2\right]\\ \label{NRam}
&+&\Lambda_S\left(\sum_{a=1}^{4}r_a^2-1\right).\ea The equations
of motion for the variables $\alpha_a(\tau)$ can be easily
integrated once and the result is \ba\label{a'm}
\dot{\alpha}_a(\tau)=\frac{C_a}{r_a^2(\tau)},\ea where $C_a$ are
arbitrary integration constants. Substituting (\ref{a'm}) back
into (\ref{NRam}), one receives an effective Lagrangian for the
four real coordinates $r_a(\tau)$\footnote{Following \cite{NR1},
we change the signs of the terms $\sim \dot{\alpha}_a^2$.} \ba\nl
\mathcal{L}&=&\frac{(4l_p\mathcal{R})^2}{4\lambda^0}\left[\sum_{a=1}^{4}
\left(\dot{r}_a^2-\frac{C_a^2}{r_a^2}\right)
-\left(8\lambda^0T_2l_p\mathcal{R}\right)^2
\sum_{a<b=1}^{4}(\omega_{a1}\omega_{b2}-\omega_{a2}\omega_{b1})^2r_a^2r_b^2
-(\kappa/2)^2\right]\\ \label{NRamr}
&+&\Lambda_S\left(\sum_{a=1}^{4}r_a^2-1\right).\ea To get
potential terms $\sim r_a^2$ instead of $\sim r_a^2r_b^2$, we use
once again the choice (\ref{c1}). In addition, we put $C_3=C_4=0$.
All this reduces the membrane Lagrangian to (after neglecting the
constant terms) \ba\label{NRml1}
L=\frac{(4l_p\mathcal{R})^2}{4\lambda^0}\sum_{a=1}^{2}\left[\dot{r}_a^2
-\left(8\lambda^0T_2l_p\mathcal{R}a\omega\right)^2
\omega_{a1}^2r_a^2-\frac{C_a^2}{r_a^2}\right]+\Lambda_S\left[\sum_{a=1}^{2}r_a^2-(1-a^2)\right],\ea
which describes Neumann-Rosochatius type integrable system. For
$C_a=0$, (\ref{NRml1}) reduces to Neumann type Lagrangian. Let us
also write down the constraints (\ref{00gf}), (\ref{0igf}) for the
present case. Actually, the third constraint $G_{02}=0$ is
satisfied identically. The other two read \ba\nl &&H\sim
\sum_{a=1}^{2}\left[\dot{r}_a^2
+\left(8\lambda^0T_2l_p\mathcal{R}a\omega\right)^2
\omega_{a1}^2r_a^2+\frac{C_a^2}{r_a^2}\right]=(\kappa/2)^2-(ab)^2,\\
\nl &&\sum_{a=1}^{2}\omega_{a1}C_a=0.\ea

Our proposal for the next type of membrane embedding is
\ba\label{4} Z_{0}=2l_p\mathcal{R}e^{i\kappa\tau},\h Z_1=0,\h
W_{a}=4l_p\mathcal{R}r_a(\sigma_i)e^{i\left[\omega_{a}\tau+\alpha_a(\sigma_i)\right]},\ea
for which the Lagrangian (\ref{geml}) reduces to \ba\label{NRmL1}
\mathcal{L}&=&-\frac{(4l_p\mathcal{R})^2}{4\lambda^0}\left\{\left(8\lambda^0T_2l_p\mathcal{R}\right)^2
\sum_{a<b=1}^{4}\left[(\p_1r_a\p_2r_b-\p_2r_a\p_1r_b)^2\right. \right. \\
\nl &+& \left. \left.
(\p_1r_a\p_2\alpha_b-\p_2r_a\p_1\alpha_b)^2r_b^2 +
(\p_1\alpha_a\p_2r_b-\p_2\alpha_a\p_1r_b)^2r_a^2\right. \right.
 \\ \nl
&+& \left. \left.(\p_1\alpha_a\p_2\alpha_b-\p_2\alpha_a\p_1\alpha_b)^2r_a^2r_b^2 \right]\right. \\
\nl &+&
\left.\sum_{a=1}^{4}\left[\left(8\lambda^0T_2l_p\mathcal{R}\right)^2(\p_1r_a\p_2\alpha_a-\p_2r_a\p_1\alpha_a)^2
-\omega_a^2\right]r_a^2 +(\kappa/2)^2\right\}\\ \nl
&+&\Lambda_S\left(\sum_{a=1}^{4}r_a^2-1\right).\ea If we restrict
ourselves to the case (\ref{smch}) and \ba\nl
\alpha_1=\alpha_1(\s_1),\h \alpha_2=\alpha_2(\s_1),\h
\alpha_3,\alpha_4=constants,\ea we obtain \ba\label{ml4}
\mathcal{L}&=&-\frac{(4l_p\mathcal{R})^2}{4\lambda^0}\left[\left(8\lambda^0T_2l_p\mathcal{R}ab\right)^2
\sum_{a=1}^{2}\left(r_a'^2+r_a^2\alpha_a'^2\right)
-\sum_{a=1}^{2}\omega_a^2r_a^2 +(\kappa/2)^2- (a\omega)^2\right]
.\ea After integrating the equations of motion for $\alpha_a$ once
and replacing the solution into (\ref{ml4}), one arrives
at\footnote{After changing the corresponding signs and ignoring
the constant terms as before.} \ba\label{nrl2}
L&=&\frac{(4l_p\mathcal{R})^2}{4\lambda^0}\sum_{a=1}^{2}\left[\left(8\lambda^0T_2l_p\mathcal{R}ab\right)^2
r_a'^2-\omega_a^2r_a^2-\left(8\lambda^0T_2l_p\mathcal{R}ab\right)^2\frac{C_a^2}{r_a^2}\right]
\\ \nl &+& \Lambda_S\left[\sum_{a=1}^{2}r_a^2-(1-a^2)\right].\ea
The above Lagrangian represents particular case of the
Neumann-Rosochatius integrable system. For $C_a=0$, (\ref{nrl2})
coincides with (\ref{nl2}). The constraints (\ref{00gf}),
(\ref{0igf}) for the case under consideration are given by \ba\nl
&&H\sim
\sum_{a=1}^{2}\left[\left(8\lambda^0T_2l_p\mathcal{R}ab\right)^2
r_a'^2+\omega_a^2r_a^2+\left(8\lambda^0T_2l_p\mathcal{R}ab\right)^2\frac{C_a^2}{r_a^2}\right]
=(\kappa/2)^2- (a\omega)^2, \\ \nl
&&\sum_{a=1}^{2}\omega_{a}C_a=0,\h G_{02}\equiv 0.\ea

Our last example of membrane embedding is connected to the spiky
strings \cite{MK,SR} and giant magnons \cite{HM} configurations on
$S^5$. It reads \ba\label{5}
&&Z_{0}=2l_p\mathcal{R}e^{i\kappa\tau},\h Z_1=0,\h
W_{a}=4l_p\mathcal{R}r_a(\xi,\eta)e^{i\left[\omega_{a}\tau+\mu_a(\xi,\eta)\right]},\\
\nl &&\xi=\alpha\sigma_1+\beta\tau,\h
\eta=\gamma\sigma_2+\delta\tau,\ea where $\alpha$, $\beta$,
$\gamma$, $\delta$ are constants. For this ansatz, the membrane
Lagrangian (\ref{geml}) takes the form ($\p_\xi=\p/\p\xi$,
$\p_\eta=\p/\p\eta$) \ba\label{NRmL2}
\mathcal{L}&=&-\frac{(4l_p\mathcal{R})^2}{4\lambda^0}
\left\{\left(8\lambda^0T_2l_p\mathcal{R}\alpha\gamma\right)^2
\sum_{a<b=1}^{4}\left[(\p_\xi r_a\p_\eta r_b-\p_\eta r_a\p_\xi r_b)^2\right. \right. \\
\nl &+& \left. \left. (\p_\xi r_a\p_\eta\mu_b-\p_\eta
r_a\p_\xi\mu_b)^2r_b^2 + (\p_\xi\mu_a\p_\eta
r_b-\p_\eta\mu_a\p_\xi r_b)^2r_a^2\right. \right.
 \\ \nl
&+& \left. \left.(\p_\xi\mu_a\p_\eta\mu_b-\p_\eta\mu_a\p_\xi\mu_b)^2r_a^2r_b^2 \right]\right. \\
\nl &+&
\left.\sum_{a=1}^{4}\left[\left(8\lambda^0T_2l_p\mathcal{R}\alpha\gamma\right)^2
(\p_\xi r_a\p_\eta\mu_a-\p_\eta r_a\p_\xi\mu_a)^2
-\omega_a^2\right]r_a^2 +(\kappa/2)^2\right\}\\ \nl
&+&\Lambda_S\left(\sum_{a=1}^{4}r_a^2-1\right).\ea Now, we choose
to consider the particular case \ba\nl &&r_{1}=r_{1}(\xi),\h
r_{2}=r_{2}(\xi),\h \omega_3=\pm\omega_4=\omega,\\ \nl
&&r_3=r_3(\eta)=a\sin(b\eta+c),\h r_4=r_4(\eta)=a\cos(b\eta+c),\h
a<1, \\ \nl &&\mu_1=\mu_1(\xi),\h \mu_2=\mu_2(\xi),\h
\mu_3,\mu_4=constants,\ea and receive (prime is used for $d/d\xi$)
\ba\nl \mathcal{L}&=&-\frac{(4l_p\mathcal{R})^2}{4\lambda^0}
\left\{\sum_{a=1}^{2}\left[(A^2-\beta^2)r_a'^2 +
(A^2-\beta^2)r_a^2\left(\mu'_a-\frac{\beta\omega_a}{A^2-\beta^2}\right)^2
- \frac{A^2}{A^2-\beta^2}\omega_a^2 r_a^2\right]\right. \\
\label{nrml2} &+&\left.
(\kappa/2)^2-a^2(\omega^2+b^2\delta^2)\right\} +
\Lambda_S\left[\sum_{a=1}^{2}r_a^2-(1-a^2)\right],\ea where \ba\nl
A^2\equiv
\left(8\lambda^0T_2l_p\mathcal{R}ab\alpha\gamma\right)^2.\ea A
single time integration of the equations of motion for $\mu_a$
following from the above Lagrangian gives \ba\label{mu'}
\mu'_a=\frac{1}{A^2-\beta^2}\left(\frac{C_a}{r_a^2}+\beta\omega_a\right).\ea
Substituting (\ref{mu'}) back into (\ref{nrml2}), one obtains the
following effective Lagrangian for the coordinates
$r_a(\xi)$\footnote{Following \cite{NR2}, we change the overall
sign, the signs of the terms $\sim C_a^2$, and discard the
constant terms.} \ba\nl L&=&\frac{(4l_p\mathcal{R})^2}{4\lambda^0}
\sum_{a=1}^{2}\left[(A^2-\beta^2)r_a'^2 -
\frac{1}{A^2-\beta^2}\frac{C_a^2}{r_a^2}
- \frac{A^2}{A^2-\beta^2}\omega_a^2 r_a^2\right] \\
\label{nrl5}
&+&\Lambda_S\left[\sum_{a=1}^{2}r_a^2-(1-a^2)\right].\ea Let us
write down the constraints (\ref{00gf}), (\ref{0igf}) for the
present case. To achieve more close correspondence with the string
on $AdS_5\times S^5$, we want the third one to be satisfied
identically. To this end, since $G_{02}\sim (ab)^2\gamma\delta,$
we set $\delta=0$, i.e. $\eta=\gamma\sigma_2$. Then, the first two
constraints give \ba\nl &&H\sim
\sum_{a=1}^{2}\left[(A^2-\beta^2)r_a'^2+
\frac{1}{A^2-\beta^2}\frac{C_a^2}{r_a^2}+
\frac{A^2}{A^2-\beta^2}\omega_a^2
r_a^2\right]=\frac{A^2+\beta^2}{A^2-\beta^2}\left[(\kappa/2)^2-(a\omega)^2\right],
\\ \nl &&\sum_{a=1}^{2}\omega_{a}C_a +
\beta\left[(\kappa/2)^2-(a\omega)^2\right]=0.\ea The Lagrangian
(\ref{nrl5}), in full analogy with the string considerations (see
(\ref{NRSM}) above or (2.26) of \cite{NR2}), corresponds to
particular case of the $n$-dimensional Neumann-Rosochatius
integrable system.

\subsection{Energy and angular momenta}
Here, we will evaluate the explicit expressions for the conserved
charges, and will obtain relations between them, for the five
membrane configurations considered above.

The energy $E$ and the angular momenta $J_a$ can be computed by
using the equalities \ba\nl E=-\int
d^2\sigma\frac{\p\mathcal{L}}{\p\kappa},\h J_a=\int
d^2\sigma\frac{\p\mathcal{L}}{\p(\p_0\varphi_a)}.\ea Then, for all
ansatzes we used, the energy is given by \ba\label{E} E=2^3(\pi
l_p\mathcal{R})^2\frac{\kappa}{\lambda^0}.\ea

For the first embedding (\ref{1}), $J_a=0$ for $a=1,2,3,4$, so the
only nontrivial conserved quantity is the membrane energy.

For the second embedding (\ref{2}), one obtains \ba\nl
J_a=2^3(l_p\mathcal{R})^2\frac{\omega_a}{\lambda^0}\int d^2\sigma
r_a^2,\h a=1,2,3,4.\ea We will consider the cases $a=1,2$ and
$a=3,4$ separately. According to (\ref{smch}),
$r_{1,2}=r_{1,2}(\sigma_{1})$, which leads to \ba\nl
J_a=\pi(4l_p\mathcal{R})^2\frac{\omega_a}{\lambda^0}\int d\sigma_1
r_a^2(\s_1),\h a=1,2.\ea Combining these two equalities with
(\ref{E}) and taking into account the constraint \ba\nl
\sum_{a=1}^{2}r_a^2-(1-a^2)=0,\ea one arrives at the energy-charge
relation \ba\nl
\frac{E}{\kappa}=\frac{1}{4(1-a^2)}\left(\frac{J_1}{\omega_1}+\frac{J_2}{\omega_2}\right).\ea
As usual, we have linear dependence $E(J_1,J_2)$ before taking the
semiclassical limit. We comment on this limit in the next section.

Now, let us turn to the case $a=3,4$. In accordance with
(\ref{smch}), we have  \ba\nl
&&J_3=\pi(4l_p\mathcal{R})^2\frac{\omega
a^2}{\lambda^0}\int_{0}^{2\pi} d\sigma_2 \sin^2(b\s_2+c),\\ \nl
&&J_4=\pm\pi(4l_p\mathcal{R})^2\frac{\omega
a^2}{\lambda^0}\int_{0}^{2\pi} d\sigma_2 \cos^2(b\s_2+c).\ea By
using the periodicity conditions \ba\nl
r_a(\s_i)=r_a(\s_i+2\pi),\ea  which imply $b=\pm 1, \pm 2,...$,
one obtains \ba\nl J_3=\pm J_4=(4\pi l_p\mathcal{R})^2\frac{\omega
a^2}{\lambda^0}.\ea In order to reproduce the string case, we can
set $\omega=0$, and thus $J_3=J_4=0$.

For the third embedding (\ref{3}), the angular momenta are given
by \ba\nl J_a=2^5(\pi l_p\mathcal{R})^2 \frac{C_a}{\lambda^0},\h
a=1,2;\h J_3=J_4=0.\ea This leads to the energy-charge relation
\ba\nl
\frac{E}{\kappa}=\frac{1}{8}\left(\frac{J_1}{C_1}+\frac{J_2}{C_2}\right).\ea

For the forth embedding (\ref{4}), the expressions for the
conserved charges and the relation between them are the same as
for the second membrane embedding (\ref{2}).

Finally, for the fifth embedding (\ref{5}), $J_3=J_4=0$ for
$\omega=0$. The other two angular momenta are \ba\nl
J_a=\frac{\pi(4l_p\mathcal{R})^2}{\lambda^0\alpha(A^2-\beta^2)}\int
d\xi \left(\beta C_a + A^2\omega_a r_a^2\right),\h a=1,2.\ea
Rewriting (\ref{E}) as\footnote{For the limits of the integrals
over $\xi$ see \cite{NR2}.} \ba\nl
E=\frac{4\pi(l_p\mathcal{R})^2\kappa}{\lambda^0\alpha}\int
d\xi,\ea we obtain the energy-charge relation \ba\nl
\frac{4}{A^2-\beta^2}\left[A^2(1-a^2) +
\beta\sum_{a=1}^{2}\frac{C_a}{\omega_a}\right]\frac{E}{\kappa}
=\sum_{a=1}^{2}\frac{J_a}{\omega_a},\ea in full analogy with the
string case. Namely, for strings on $AdS_5\times S^5$, the result
in conformal gauge is \cite{NR2} \ba\nl
\frac{1}{\alpha^2-\beta^2}\left(\alpha^2+
\beta\sum_{a}\frac{C_a}{\omega_a}\right)\frac{E}{\kappa}
=\sum_{a}\frac{J_a}{\omega_a}.\ea

Concluding this section, let us make two remarks.

It may seems that the membrane configurations considered here are
chosen randomly. However, they correspond exactly to {\it all}
string embeddings in the $R\times S^5$ subspace of $AdS_5\times
S^5$ solution of type IIB string theory, which are known to lead
the Neumann and Neumann-Rosochatius dynamical systems \cite{N},
\cite{NR1}, \cite{NR2}.

Let us also note that our starting ansatz is a particular case of
(2.9) in \cite{BRR04}.

\setcounter{equation}{0}
\section{Concluding remarks}
We have found here several types of membrane embedding into the
$AdS_4\times S^7$ background, which are related to the Neumann and
Neumann-Rosochatius integrable systems, thus reproducing from
M-theory viewpoint part of the results established for strings on
$AdS_5\times S^5$. In particular, our Lagrangian (\ref{nrl5}),
being completely analogous to the one given in (2.26) of
\cite{NR2}, should lead to the same energy-charge relation for the
giant magnon solution with two angular momenta (see also
\cite{BR06}, \cite{B06}). Moreover, the single spike solutions of
\cite{IK0705} can be reproduced also from membranes on
$AdS_4\times S^7$ \cite{PBRR}. In addition, one can consider the
correspondence between the Neumann and Neumann-Rosochatius
integrable systems arising from membranes and the continuous limit
of integrable spin chains at the level of actions, as is done in
\cite{0706.1443}.

Besides, it is interesting to clarify the relationship with the
constant radii solutions for membranes on $AdS_4\times S^7$ found
in \cite{BRR04}. It turns out that part of them are particular
solutions of a Neumann-Rosochatius system. Let us explain this in
more detail. Consider the membrane embedding (\ref{3}), which in
view of (\ref{a'm}), reduces the Lagrangian (\ref{NRam}) to
(\ref{NRamr}). By choosing \ba\nl
\omega_{12}=\omega_{22}=\omega_{31}=\omega_{41}=0\ea as in
(\ref{c1}), one obtains \ba\nl
\sum_{a<b=1}^{4}(\omega_{a1}\omega_{b2}-\omega_{a2}\omega_{b1})^2r_a^2r_b^2
=(\omega_{11}^2r_1^2+\omega_{21}^2r_2^2)(\omega_{32}^2r_3^2+\omega_{42}^2r_4^2).\ea
For $r_{3,4}=constants$, $C_3=C_4=0$, this leads to a
Neumann-Rosochatius Lagrangian of the type (\ref{NRml1}) \ba\nl
L=\frac{(4l_p\mathcal{R})^2}{4\lambda^0}\sum_{a=1}^{2}\left[\dot{r}_a^2
-\left(8\lambda^0T_2l_p\mathcal{R}\omega_r\right)^2
\omega_{a1}^2r_a^2-\frac{C_a^2}{r_a^2}\right]+\Lambda_S\left[\sum_{a=1}^{2}r_a^2-(1-r_3^2-r_4^2)\right],\ea
where \ba\nl \omega_r^2=\omega_{32}^2r_3^2+\omega_{42}^2r_4^2.\ea
Now, let us impose the conditions \ba\nl
\alpha_a(\tau)=\omega_{a0}\tau,\h \omega_{a0}=constants,\ea as is
the case in \cite{BRR04}. These are compatible with (\ref{a'm})
for $r_a=constants$ only, i.e. for the constant radii solutions of
\cite{BRR04}, when the equations of motion and constraints reduce
to relations between the free parameters of the membrane
embedding. Therefore, the membrane solutions described in section
4 of \cite{BRR04} are solutions of a Neumann-Rosochatius system
for particular choice of the parameters $\omega_{ai}$ and $C_a$.
For $\omega_{ai}$ - arbitrary, they are solutions of a more
general system, given by the Lagrangian (\ref{NRamr}).

According to AdS-CFT correspondence, strings on $AdS_5\times S^5$
and membranes on $AdS_4\times S^7$ are dual to {\it different}
gauge theories. Hence, one is tempting to conjecture that there
should exist common integrable sectors on the field theory side.

We expect that in the framework of our approach, one can find
relations between membranes in $AdS_7\times S^4$ and Neumann and
Neumann-Rosochatius like integrable systems with indefinite
signature, analogous to (\ref{Nl}) and (\ref{NRL}).

On the other hand, we observed that only a small class of membrane
configurations described by the embedding (\ref{gme}) are captured
by the Neumann and Neumann-Rosochatius dynamical systems.
Actually, these configurations are {\it exceptional}, taking into
account the Lagrangians (\ref{NmL1}), (\ref{NmL2}), (\ref{NRam}),
(\ref{NRmL1}) and (\ref{NRmL2}). The conclusion is that there
exist many possibilities for discovering, known or new, integrable
systems dual to the membranes in M-theory.

\vspace*{.5cm}

{\bf Acknowledgments} \vspace*{.2cm}

This work is supported by NSFB grants $F-1412/04$ and
$VU-F-201/06$.



\begin{thebibliography}{}
\bibitem{M97} Juan M. Maldacena, {\it The Large N Limit of Superconformal Field Theories and
Supergravity}, Adv. Theor. Math. Phys. {\bf 2} (1998) 231-252;
Int. J. Theor. Phys. {\bf 38} (1999) 1113-1133,
[arXiv:hep-th/9711200v3].
\bibitem{GKP98} S.S. Gubser, I.R. Klebanov, A.M. Polyakov,
{\it Gauge Theory Correlators from Non-Critical String Theory},
Phys. Lett. {\bf B 428} (1998) 105-114, [arXiv:hep-th/9802109v2].
\bibitem{W98}  Edward Witten, {\it Anti De Sitter Space And
Holography}, Adv. Theor. Math. Phys. {\bf 2} (1998) 253-291,
[arXiv:hep-th/9802150v2].
\bibitem{BMN02}  David Berenstein, Juan Maldacena, Horatiu
Nastase, {\it Strings in flat space and pp waves from ${\cal N}=4$
Super Yang Mills}, JHEP 0204 (2002) 013, [arXiv:hep-th/0202021v3].
\bibitem{GKP02}S. S. Gubser, I. R. Klebanov, A. M. Polyakov,
\textit{A semi-classical limit of the gauge/string
correspondence}, Nucl. Phys. \textbf{B 636} (2002) 99-114,
[arXiv:hep-th/0204051v3].
\bibitem{MZ} J. A. Minahan and K. Zarembo,
\textit{The Bethe-ansatz for $\mathcal{N}$ = 4 super Yang-Mills},
JHEP 0303 (2003) 013, [arXiv:hep-th/0212208v3].
\bibitem{N} G. Arutyunov, S. Frolov, J. Russo, A.A. Tseytlin, {\it Spinning strings in $AdS_5\times S^5$ and integrable
systems}, Nucl. Phys. {\bf B 671} (2003) 3-50
[arXiv:hep-th/0307191v3].
\bibitem{NR1} G. Arutyunov, J. Russo, A.A. Tseytlin, {\it Spinning strings in $AdS_5\times S^5$: new integrable system
relations}, Phys. Rev. {\bf D 69} (2004) 086009
[arXiv:hep-th/0311004v2].
\bibitem{NR2} M. Kruczenski, J. Russo, A.A. Tseytlin, {\it Spiky strings and giant magnons on
$S^5$}, JHEP 0610 (2006) 002 [arXiv:hep-th/0607044v3].
\bibitem{Neumann} C. Neumann, {\it De problemate quodam mechanico, quod
ad primam integralium ultraellipticorum classem revocatur}, J.
Reine Angew. Math. {\bf 56}, (1859) 46-63; \\
O.~Babelon and M.~Talon, {\it Separation Of Variables For The
Classical And Quantum Neumann Model}, Nucl.\ Phys.\ B {\bf 379},
321 (1992)[arXiv:hep-th/9201035v1].
\bibitem {Per} A.M. Perelomov, {\it Integrable Systems of Classical Mechamics and Lie
Algebras}, Springer Verlag, 1990.
\bibitem{NR} E. Rosochatius, {\it Uber Bewegungen eines Punktes},
Dissertation at Univ. G\"otingen, Druck von Gebr. Unger, Berlin
1877; \\
J. Moser, {\it Various aspects of integrable Hamiltonian Systems},
in "Dynamical systems", Progress in Mathematics {\bf 8}, C.I.M.E.
Lectures, Bressanone, Italy, (1978) J. Coates, S. Helgason, Eds.;\\
J. Moser, {\it Geometry of Quadrics},
Chern Symposium, Berkeley (1979), p.147;\\
T.S. Ratiu, {\it The Lie algebraic interpretation of the complete
integrability of the Rosochatius system}, AIP Proceed. 88, Amer.
Inst. Physics, New York (1982), pp. 109;\\
L.~Gagnon, J.~P.~Harnad, J.~Hurtubise and P.~Winternitz, {\it
Abelian Integrals And The Reduction Method For An Integrable
Hamiltonian System}, J.\ Math.\ Phys.\  {\bf 26}, 1605 (1985);\\
R.~Kubo, W.~Ogura, T.~Saito and Y.~Yasui, {\it Geodesic flows for
the Neumann-Rosochatius systems}, YITP-97-46,
arXiv:physics/9710016v1;\\
C. Bartocci, G. Falqui and M. Pedroni, {\it A geometrical approach
to the separability of the Neumann-Rosochatius system},
arXiv:nlin.SI/0307021v1.
\bibitem{hep-th/0304255} S. Frolov, A.A. Tseytlin, {\it Multi-spin string solutions in
$AdS_5\times S^5$}, Nucl. Phys. {\bf B 668} (2003) 77-110
[arXiv:hep-th/0304255v3].
\bibitem{hep-th/0306143} S. Frolov, A.A. Tseytlin, {\it Rotating string solutions:
AdS/CFT duality in non-supersymmetric sectors}, Phys. Lett. {\bf B
570} (2003) 96-104 [arXiv:hep-th/0306143v2].
\bibitem{hep-th/0204226} S. Frolov, A.A. Tseytlin, {\it Semiclassical quantization of
rotating superstring in $AdS_5\times S^5$}, JHEP 0206 (2002) 007
[arXiv:hep-th/0204226v5].
\bibitem{hep-th/0306139} N. Beisert, J. A. Minahan, M. Staudacher, K.
Zarembo, {\it Stringing Spins and Spinning Strings }, JHEP 0309
(2003) 010 [arXiv:hep-th/0306139v2].
\bibitem{MK} M. Kruczenski, {\it Spiky strings and single trace operators in gauge
theories}, JHEP 0508 (2005) 014 [arXiv:hep-th/0410226v2].
\bibitem{SR} S. Ryang, {\it Wound and Rotating Strings in
$AdS_5\times S^5$}, JHEP 0508 (2005) 047 [arXiv:hep-th/0503239v1].
\bibitem{HM} Diego M. Hofman, Juan Maldacena, \textit{Giant
Magnons},  J. Phys. A 39 (2006) 13095-13118
[arXiv:hep-th/0604135v2].
\bibitem{Dorey1} Nick Dorey, \textit{Magnon bound states and the AdS/CFT
correspondence}, J. Phys. A 39 (2006) 13119-13128,
[arXiv:hep-th/0604175v2].
\bibitem{Dorey2}  Heng-Yu Chen, Nick Dorey, Keisuke Okamura
, {\it Dyonic Giant Magnons}, JHEP 0609 (2006) 024,
[arXiv:hep-th/0605155v2].
\bibitem{AFZ} Gleb Arutyunov, Sergey Frolov, Marija Zamaklar,
\textit{Finite-size Effects from Giant Magnons},
doi:10.1016/j.nuclphysb.2006.12.026 [arXiv:hep-th/0606126v2].
\bibitem{MTT} J.A. Minahan, A. Tirziu, A.A. Tseytlin,
\textit{Infinite spin limit of semiclassical string states}, JHEP
0608 (2006) 049, [arXiv:hep-th/0606145v2].
\bibitem{CGK06} Chong-Sun Chu, George Georgiou, Valentin V. Khoze,
{\it Magnons, Classical Strings and beta-Deformations}, JHEP 0611
(2006) 093, [arXiv:hep-th/0606220v2].
\bibitem{SV}Marcus Spradlin, Anastasia Volovich,\textit{"Dressing the Giant Magnon"}, JHEP 0610 (2006) 012
, [arXiv:hep-th/0607009v3]; Chrysostomos Kalousios, Marcus
Spradlin, Anastasia Volovich, {\it Dressing the Giant Magnon II},
arXiv:hep-th/0611033v1.
\bibitem{BobR06} N.P. Bobev, R.C. Rashkov, {\it Multispin Giant
Magnons}, Phys. Rev. {\bf D 74} (2006) 046011,
[arXiv:hep-th/0607018v3].
\bibitem{BR06} P. Bozhilov, R.C. Rashkov, {\it Magnon-like dispersion relation from
M-theory}, Nucl. Phys. {\bf B 768} [PM] (2007) 193-208
[arXiv:hep-th/0607116v3].
\bibitem{WHH06} Wung-Hong Huang, {\it Giant Magnons under NS-NS and Melvin
Fields}, JHEP 0612 (2006) 040, [arXiv:hep-th/0607161v4].
\bibitem{OS} Keisuke Okamura, Ryo Suzuki, {\it A Perspective on Classical Strings from
Complex Sine-Gordon Solitons}, Phys. Rev. {\bf D 75} (2007) 046001
[arXiv:hep-th/0609026v3].
\bibitem{H} Shinji Hirano, {\it Fat Magnon}, arXiv:hep-th/0610027v4.
\bibitem{R} Shijong Ryang, {\it Three-Spin Giant Magnons in
$AdS_5xS^5$}, JHEP 0612 (2006) 043 [arXiv:hep-th/0610037v1].
\bibitem{CDO} Heng-Yu Chen, Nick Dorey, Keisuke Okamura,
{\it The Asymptotic Spectrum of the N=4 Super Yang-Mills Spin
Chain}, arXiv:hep-th/0610295v1.
\bibitem{MS} Juan Maldacena, Ian Swanson, {\it Connecting giant magnons to the pp-wave:
An interpolating limit of $AdS_5 \times S^5$},
arXiv:hep-th/0612079v3.
\bibitem{B06} P. Bozhilov, {\it A note on two-spin magnon-like energy-charge relations
from M-theory viewpoint}, arXiv:hep-th/0612175v1.
\bibitem{M07} J. A. Minahan, {\it  Zero modes for the giant
magnon}, JHEP 0702 (2007) 048 [arXiv:hep-th/0701005v3].
\bibitem{AFGS07} Davide Astolfi, Valentina Forini, Gianluca Grignani, Gordon W.
Semenoff, {\it Gauge invariant finite size spectrum of the giant
magnon}, arXiv:hep-th/0702043v3.
\bibitem{V07} Benoit Vicedo, {\it Giant Magnons and Singular Curves},
arXiv:hep-th/0703180v1.
\bibitem{KNP07} J. Kluson, Rashmi R. Nayak, Kamal L. Panigrahi,
{\it Giant Magnon in NS5-brane Background},
arXiv:hep-th/0703244v2.
\bibitem{PS07} Georgios Papathanasiou, Marcus Spradlin,
{\it Semiclassical Quantization of the Giant Magnon},
arXiv:0704.2389v1 [hep-th].
\bibitem{BRR04} J. Brugues, J. Rojo and J. G. Russo, {\it Non-perturbative states in type II
superstring theory from classical spinning membranes}, Nucl. Phys.
{\bf B 710} (2005) 117-138, arXiv:hep-th/0408174v2.
\bibitem{IK0705}  Riei Ishizeki, Martin Kruczenski, {\it Single spike solutions for strings on S2 and
S3}, arXiv:0705.2429v1 [hep-th].
\bibitem{PBRR} P. Bozhilov, R.C. Rashkov, in preparation.
\bibitem {0706.1443} P. Bozhilov, {\it Spin chain from membrane and the
Neumann-Rosochatius integrable system}, arXiv:0706.1443v2
[hep-th].

\end{thebibliography}
\end{document}